\documentclass[usenatbib]{mn2e}

\usepackage{amssymb}
\usepackage{amsmath}
\usepackage{epsfig}
\title{Evolutionary Tracks of Individual Quasars in the Mass-Luminosity Plane}
\author[Charles L. Steinhardt, Martin Elvis, and Mihail Amarie]
       {Charles L. Steinhardt$^{1,2}$, Martin Elvis$^{2}$, and Mihail Amarie$^{3}$  \\ 
$^{1}$Institute for the Physics and Mathematics of the Universe, University of Tokyo, Kashiwanoha 5-1-5, Kashiwa, Chiba, Japan\\
$^{2}$Harvard-Smithsonian Center for Astrophysics, 60 Garden St, Cambridge, MA 02138 \\
$^{3}$Department of Physics, Jadwin Hall, Princeton, NJ 08544}
\date{\today}
\begin{document}

\let\la=\lesssim     
\let\ga=\gtrsim
\def\case#1#2{\hbox{$\frac{#1}{#2}$}}
\def\slantfrac#1#2{\hbox{$\,^#1\!/_#2$}}
\def\onehalf{\slantfrac{1}{2}}
\def\onethird{\slantfrac{1}{3}}
\def\twothirds{\slantfrac{2}{3}}
\def\onequarter{\slantfrac{1}{4}}
\def\threequarters{\slantfrac{3}{4}}
\def\ubvr{\hbox{$U\!BV\!R$}}            
\def\ub{\hbox{$U\!-\!B$}}               
\def\bv{\hbox{$B\!-\!V$}}               
\def\vr{\hbox{$V\!-\!R$}}               
\def\ur{\hbox{$U\!-\!R$}}
\def\ion#1#2{#1$\;${\small\rm\@Roman{#2}}\relax}

\def\aj{\rm{AJ}}                    
\def\apj{\rm{ApJ}}                 
\def\apjl{\rm{ApJ}}                
\def\apjs{\rm{ApJS}}                        
\def\mnras{\rm{MNRAS}}

\maketitle

\label{firstpage}

\begin{abstract}
Previous work on the quasar mass-luminosity plane indicates the possibility that quasars of the same central black hole mass might follow a common evolutionary track, independent of the properties of the host galaxy.  We consider two simple models for the evolution of individual quasars. Requiring these tracks to lie within the observed quasar locus at all redshifts strongly constrains the model parameters, but does allow some solutions.  These solutions include a family of tracks with similar shape but different initial masses that might match the observed quasar distributions at all redshifts $z < 2.0$.  This family of solutions is characterized by short (1-2 Gyr) lifetimes, a duty cycle in which the quasar is on at least 25\% of the time, and a rapid decline in Eddington ratio, perhaps with $L_{Edd} \propto t^{-6}$ or steeper.
\end{abstract}

\begin{keywords}
black hole physics --- galaxies: evolution --- galaxies: nuclei --- quasars:
general --- accretion, accretion discs
\end{keywords}

\section{Introduction}
\label{sec:trintro}
Supermassive black holes (SMBH), are found at the centers of nearly every galaxy where there have been sensitive searches.  These SMBH are the end products of a long lifecycle comprising a seeding mechanism, a growth stage and, finally, the present state with nearly zero growth.  At various points during this cycle, individual SMBH are viewed in many states, including as Type 1 and Type 2 quasars, Seyfert galaxies, and quiescence.  The Soltan argument \citep{Soltan1982} indicates that most of the final SMBH mass is gained through luminous accretion while the SMBH and its surrounding region are in a quasar phase.  It has recently been demonstrated that quasar masses and luminosities are more strongly constrained at each redshift than expected from simple models \citep{Steinhardt2010a,Steinhardt2010b}.  In this work, we use these constraints as a probe of SMBH accretion tracks to produce limits on possible feedback mechanisms.

These constraints include new boundaries on quasar luminosities on both the bright and faint ends at $0.8 < z < 4.1$.  Several examples of quasar distributions in the mass-luminosity plane are shown in \S~\ref{sec:tracks}, and on the high-mass end, the entire quasar distribution lies above the Sloan Digital Sky Survey (SDSS) detection limit, bounded on the bright end by a sub-Eddington boundary \citep{Steinhardt2010a} and on the faint end by high-mass, low-luminosity boundary \citep{Steinhardt2010b}.  At these redshifts, the entire range of quasar luminosities for a given mass spans less than 1 dex everywhere that it can be measured \citep{Steinhardt2010b}.   Given the $\sim 0.3$ dex variability of quasars \citep{structurefunction}, the intrinsic width of this distribution must be quite narrow.  For fixed mass and redshift, the luminosity distribution is sharply peaked at one, characteristic luminosity.  This luminosity drops significantly below the Eddington limit for larger central black hole masses at each redshift.  The lowest-mass quasars at each redshift can typically approach their Eddington luminosity, while the highest-mass quasars at each redshift typically are below $\sim 10\%$ of Eddington.  At $0.2 < z < 0.8$, the SDSS detection limit is within 1 dex of the bright-end, sub-Eddington boundary, so we cannot test whether the width of the entire luminosity range is less than 1 dex.  However, there is again a sharp peak at one, characteristic luminosity and we might expect that with improved detection, quasars at $0.2 < z < 0.8$ would exhibit identical properties to those at higher redshifts.  The observed characteristic luminosity evolves with mass at fixed redshift and evolves with redshift at fixed mass.  

The remarkable implication of these constraints is that there is a characteristic luminosity for quasar accretion dependent only upon the central SMBH mass and redshift.  In particular, even though we would expect a large sample of supermassive black holes of a given mass and redshift to lie within host galaxies of a range of different virialization times, morphologies, star formation rates and merger histories, the corresponding quasars have a common characteristic luminosity.  It is also unexpected that the characteristic luminosity evolves with redshift but also remains narrow at fixed $z$.  This combination appears to require some non-trivial evolution of accretion rates that occurs simultaneously in different host galaxies.  A discussion of physical models capable of producing this behavior is left for future work.  In this paper, we use the evolution of these narrow luminosity ranges to constrain quasar accretion histories, i.e., tracks taken in the mass-luminosity plane.

 Recent work has investigated the possibility that virial mass estimates may contain a mass-dependent bias \citep{Onken2008,Risaliti2009,Rafiee2010}.  These proposed mass-dependent biases all have the property that black holes measured in \cite{Shen2008} to have similar masses also have similar masses after the corrections are applied (and black holes measured to have dissimilar masses have dissimilar masses after correction).  As a result, the luminosity distribution at fixed mass and redshift continues to be equally narrow after correcting for any of these biases, and evolutionary tracks will yield comparably restrictive, though different, bounds to those described in this work.

A bias that takes the true $M-L$ distribution and maps it into one that fills less of the estimated $M-L$ parameter space will affect the tracks that one derives (and perhaps the validity of the method for a strong enough bias).  This method would produce less restrictive bounds under a correction moving objects measured to have substantially different black hole masses in \citet{Shen2008} to the same corrected masses, and thus widening the luminosity distribution at fixed $M$ and $z$.  

The quasars with lowest central black hole mass at each redshift can approach their Eddington luminosity.  As they accrete, their mass increases, and as quasars move both to higher masses and later times, their characteristic Eddington ratio decreases.  The high-mass boundary is also sharp and evolving with redshift \citep{Steinhardt2010b}, so that most SMBH survive as quasars until they reach this boundary.  The sharp, characteristic luminosity at fixed mass and redshift, and its seeming independence of host galaxy parameters, imply that these surviving quasars take similar paths in mass and luminosity as a function of time from when they first appear near $L_{Edd}$ until turnoff.  

While for each individual quasar there is only one snapshot available showing its mass and luminosity at one redshift, we can reconstruct possible histories for that quasar by requiring that its evolution would, at each redshift, produce a mass and luminosity lying within the observed quasar locus in the $M-L$ plane, turning off when the quasar has become one of the highest-mass SMBH at some redshift.  We find that the specific path taken is sensitive to the SMBH accretion history, and is sensitive to feedback between the SMBH and its environment.  In \S~\ref{sec:tracks}, we develop two simple, parametrized models for SMBH accretion history following \citet{Hopkins2009} and show that the resulting quasar tracks tightly constrain their input parameters.  In \S~\ref{sec:parameters}, we fit these models to the SDSS quasar mass-luminosity plane and consider the properties that seem to be required of these tracks.  We discuss the implications of these results in \S~\ref{sec:trdiscussion}.

\section{Mass-Luminosity Plane Boundaries}
\label{sec:mlboundaries}

As shown in Papers I \citep{Steinhardt2010a} and II \citep{Steinhardt2010b}, quasars in the SDSS DR5 catalog \citep{SDSSDR5} are bounded at every redshift $z < 4.1$ (well-defined at $z < 2.0$) by a sub-Eddington boundary (SEB) on the bright end and by a high-mass, low-luminosity boundary (HMLLB) on the faint end of the quasar sample.  These two boundaries restrict the observed luminosities for quasars at fixed mass and redshift to a range spanning $\sim 1$ dex.  These boundaries are defined by 95\% dropoff in number density from the peak number density at each mass \citep{Steinhardt2010a}.  The location of this range evolves with both mass and redshift, lying within the boundaries given in Table \ref{table:boundaries}.  Because H$\beta$-based and Mg{\small II}-based virial mass estimators appear to be better than C{\small IV}-based estimators \citep{Shen2008,Marconi2009,Steinhardt2010c}, we will restrict ourselves these two estimators, and therefore to $0.2 < z < 2.0$, when producing individual quasar tracks.
\begin{table}
\caption{Locations of the sub-Eddington boundaries (SEB) and high-mass, low-luminosity boundaries (HMLLB) at $0.2 < z < 2.0$, where $M \equiv \log M_{BH}/M_\odot$.}
\begin{center}
\begin{tabular}{|c|c|c|}
\hline 
$z$ & SEB ($\log L_{\max}$) & HMLLB ($\log L_{\min}$) \\
\hline 
H$\beta$ & & \\
\hline
0.2-0.4 & 0.37 M + 42.66 & detection-limited \\
0.4-0.6 & 0.45 M + 42.17 & detection-limited \\
0.6-0.8 & 0.61 M + 41.14 & detection-limited \\
\hline
Mg{\small II} & & \\
\hline
0.8-1.0 & 0.67 M + 40.66 & 0.54 M + 40.97 \\
1.0-1.2 & 0.67 M + 40.78 & 0.35 M + 42.77 \\
1.2-1.4 & 0.73 M + 40.30 & 0.44 M + 42.01 \\
1.4-1.6 & 0.68 M + 40.73 & 0.47 M + 41.80 \\
1.6-1.8 & 0.51 M + 42.40 & 0.44 M + 42.20 \\
1.8-2.0 & 0.42 M + 43.24 & 0.31 M + 43.50 \\
\hline  
\end{tabular}
\end{center}
\label{table:boundaries}
\end{table}

For each of the quasars in the SDSS catalog, an accompanying optical spectrum allows a determination of virial SMBH mass, bolometric luminosity, and redshift.  Each SDSS quasar is one snapshot of the evolution of a particular SMBH along some track in mass, luminosity, and time.  The parameters describing that track describe the accretion history of the SMBH and reflect the feedback mechanism between the SMBH and its surrounding environment.  Because the final mass and path in mass, luminosity, and time taken by each model SMBH is sensitive to all four parameters, successful model quasar tracks would correspond to a highly specific accretion history and therefore inform us about the nature of AGN-galaxy feedback.

\section{Quasar Accretion Histories: Tracks in the Mass-Luminosity Plane}
\label{sec:tracks}

We consider two models for the decline in Eddington ratio $L/L_{Edd}$ as a function of time: (1) an exponential decay $L/L_{Edd} = e^{-(t-t_0)/k_e}$ and (2) a power-law decay $L/L_{Edd} = (t-t_0)^{-k_p}$.  The first model is inspired by arguments in \citep{Hopkins2009}, while the second uses a simple alternative form to give an indication of the dependence on the assumed function.  Both models are four-parameter fits, requiring a mass $M_0$ and time $t_0$ where the quasar departs from its Eddington luminosity, a decay slope $k$, and an accretion conversion factor $\kappa = L/\dot{M} = \epsilon/l$, the ratio of the radiative efficiency $\epsilon$ to the fraction of its lifetime $l$ that the quasar is turned on (its duty cycle).  If the quasar is always turned on, the growth rate will be the instantaneous $\kappa = L/\dot{M} = \epsilon$  Our models make the simplifying assumption that the duty cycle and radiative efficiency are not time- or mass-dependent.  Observations constraining quasar duty cycles, or a theoretical model constraining radiative efficiency, could allow this degeneracy to be broken.

Existing observations and theory restrict the range of accretion conversion factors $\kappa$.  We can estimate $l$ from the observed fraction of galaxies hosting quasars $f_Q$.  At low redshift $f_Q < 0.05$, and it increases to approach $f_Q \sim 0.2$ near $z=2$ (``quasar evolution'', \citet{Martini2009}).  Some SMBH might either not yet have reached a quasar state or have already turned off at the time $f_Q$ is calculated.  

The observed fraction of quasars $f_Q$ is the product of $l$ and the fraction of galaxies going through a quasar phase at that redshift.  Therefore, a measurement of $f_Q = 0.2$ places a lower bound of $0.2$ on $l$ at $z=2$, but $l$ might be considerably larger if relatively few galaxies contain a SMBH in a luminous accretion phase at that redshift.  If only 20\% of galaxies were capable of producing a quasar at that redshift, quasars might be permanently on from the start of their quasar phase until turnoff.  Theoretical models of accretion disks produce radiative efficiencies in the 8-20\% range, with 10\% perhaps favored \citep{Yu2002}.  Combining these limits gives $0.08 < \kappa < 1$, with the preferred values of $l = 0.2$ and $\epsilon = 0.1$ yielding a preferred value of $\kappa = 0.5$.  

\begin{figure}
  \epsfxsize=3in\epsfbox{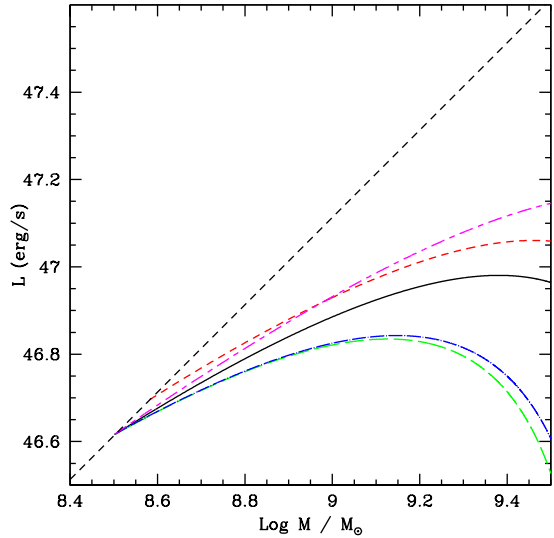}
\caption{Black: Evolution in the mass-luminosity plane of an individual quasar from $10^{8.5} M_\odot$, starting at the Eddington luminosity and declining in Eddington ratio as $L/L_{Edd} = (t-t_0)^{-6}$, or $k_p = 6$.  20\% increases in the initial mass (red), initial time (magenta), power-law slope (green), and conversion factor $\kappa$ (blue) are also indicated.  The Eddington luminosity is shown as the black, dashed line.}
\label{fig:qsomodelml}
\end{figure}

For each set of model parameters, we construct the path in the $M-L$ plane taken by one quasar obeying that model.  Figure \ref{fig:qsomodelml} shows the track taken by an individual quasar starting at $10^{8.5} M_\odot$ and $L_{Edd}$ and declining in Eddington ratio as a power law $L/L_{Edd} = (t-t_0)^{-6}$ with $\kappa = 0.5$ (black, solid line).  
20\% increases in the initial mass (red), initial time (magenta), power-law slope (green), or $\kappa$ (blue) produce different luminosities at higher mass, spanning $\sim 0.6$ dex in total.  The luminosity is most sensitive to changes in the power-law slope $k$ and conversion factor $\kappa$.  Changes in $k$ and $\kappa$ are nearly degenerate (steep values of $k$ are preferred, as in \S~\ref{sec:parameters}).

\begin{figure}
  \epsfxsize=3in\epsfbox{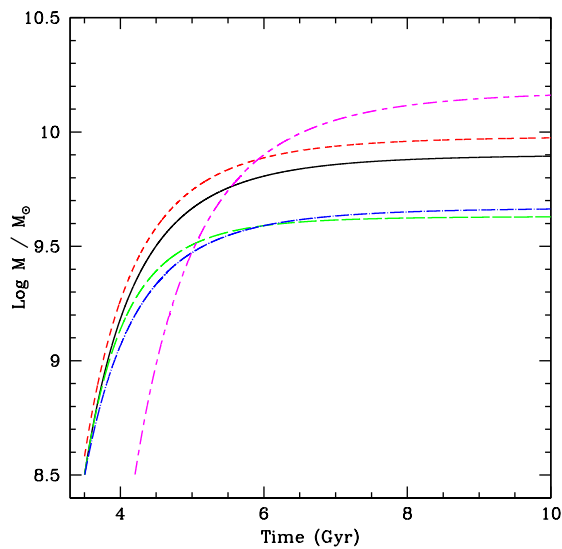}
\caption{Black: Evolution in the time-mass plane of an individual quasar from $10^{8.5} M_\odot$, starting at the Eddington luminosity and declining in Eddington ratio as $L/L_{Edd} = (t-t_0)^{-6}$.  20\% increases in the initial mass (red), initial time (magenta), power-law slope (green), and conversion factor $\kappa$ (blue) are also indicated.}
\label{fig:qsomodeltm}
\end{figure}

\begin{figure}
  \epsfxsize=3in\epsfbox{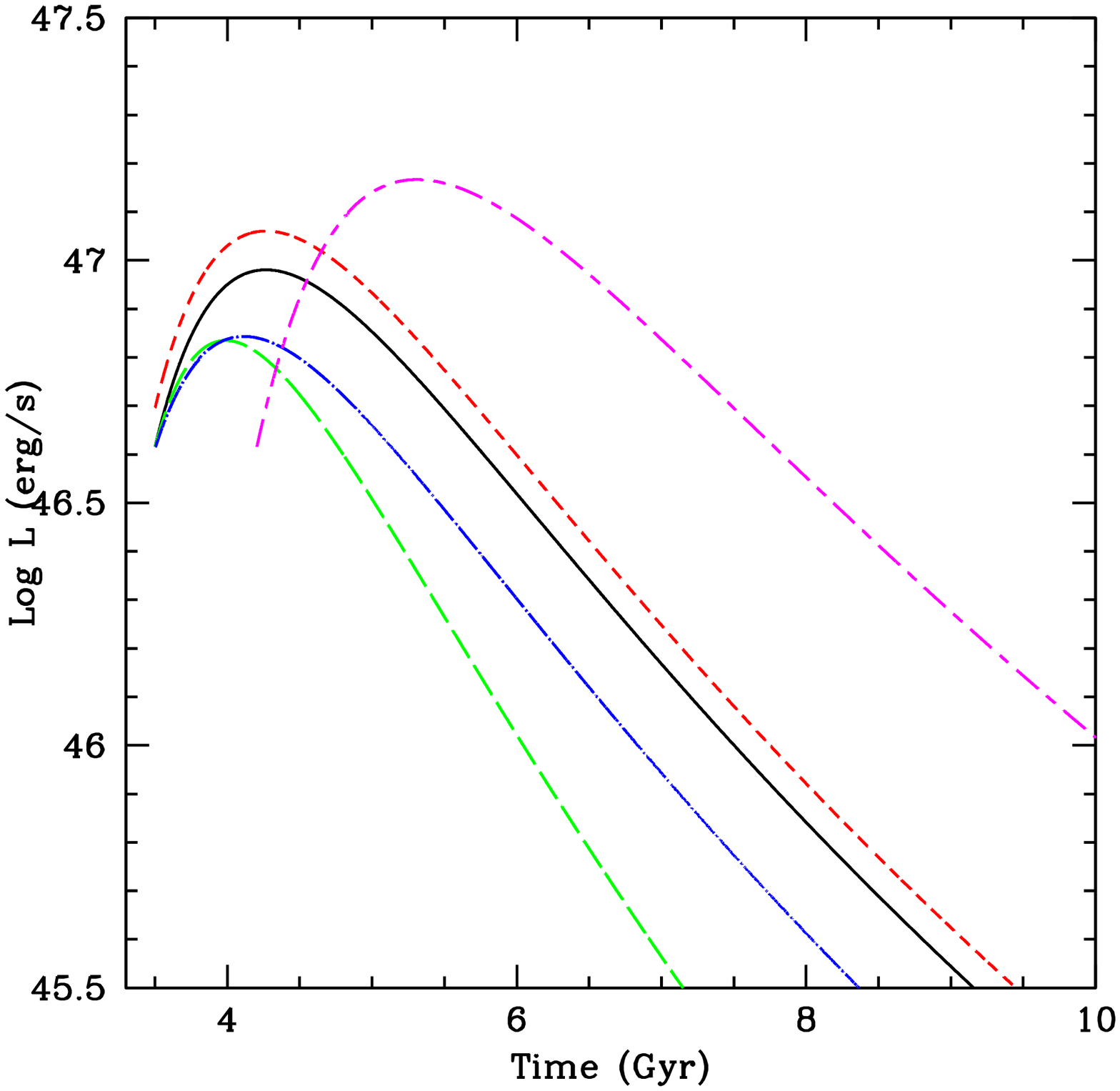}
\caption{Black: Evolution in the time-luminosity plane of an individual quasar from $10^{8.5} M_\odot$, starting at the Eddington luminosity and declining in Eddington ratio as $L/L_{Edd} = (t-t_0)^{-6}$.  20\% increases in the initial mass (red), initial time (magenta), power-law slope (green), and conversion factor $\kappa$ (blue) are also indicated.  Note that although the Eddington ratio is strongly decreasing, the increasing black hole mass results in a luminosity that first increases, then decreases.}
\label{fig:qsomodeltl}
\end{figure}

\begin{figure}
  \epsfxsize=3in\epsfbox{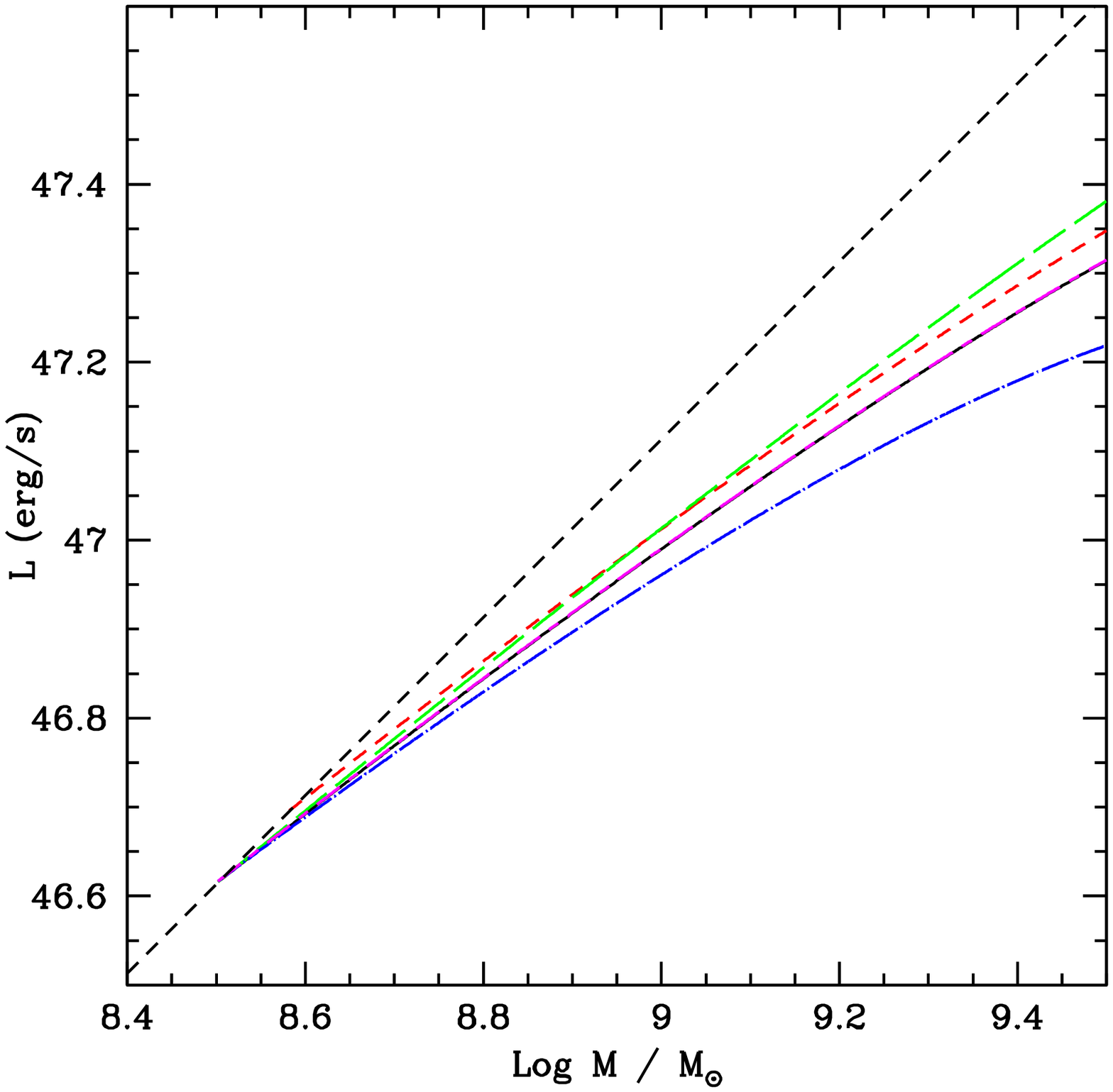}
\caption{Black: Evolution in the mass-luminosity plane of an individual quasar from $10^{8.5} M_\odot$, starting at the Eddington luminosity and declining in Eddington ratio as $L/L_{Edd} = e^{-(t-t_0)/1.0}$, or $k_e = 1$ Gyr.  20\% increases in the initial mass (red), initial time (magenta), power-law slope $k_p$ (green), and conversion factor $\kappa$ (blue) are also indicated.  The Eddington luminosity is shown as the black, dashed line.}
\label{fig:qsomodelmlexp}
\end{figure}

\begin{figure}
  \epsfxsize=3in\epsfbox{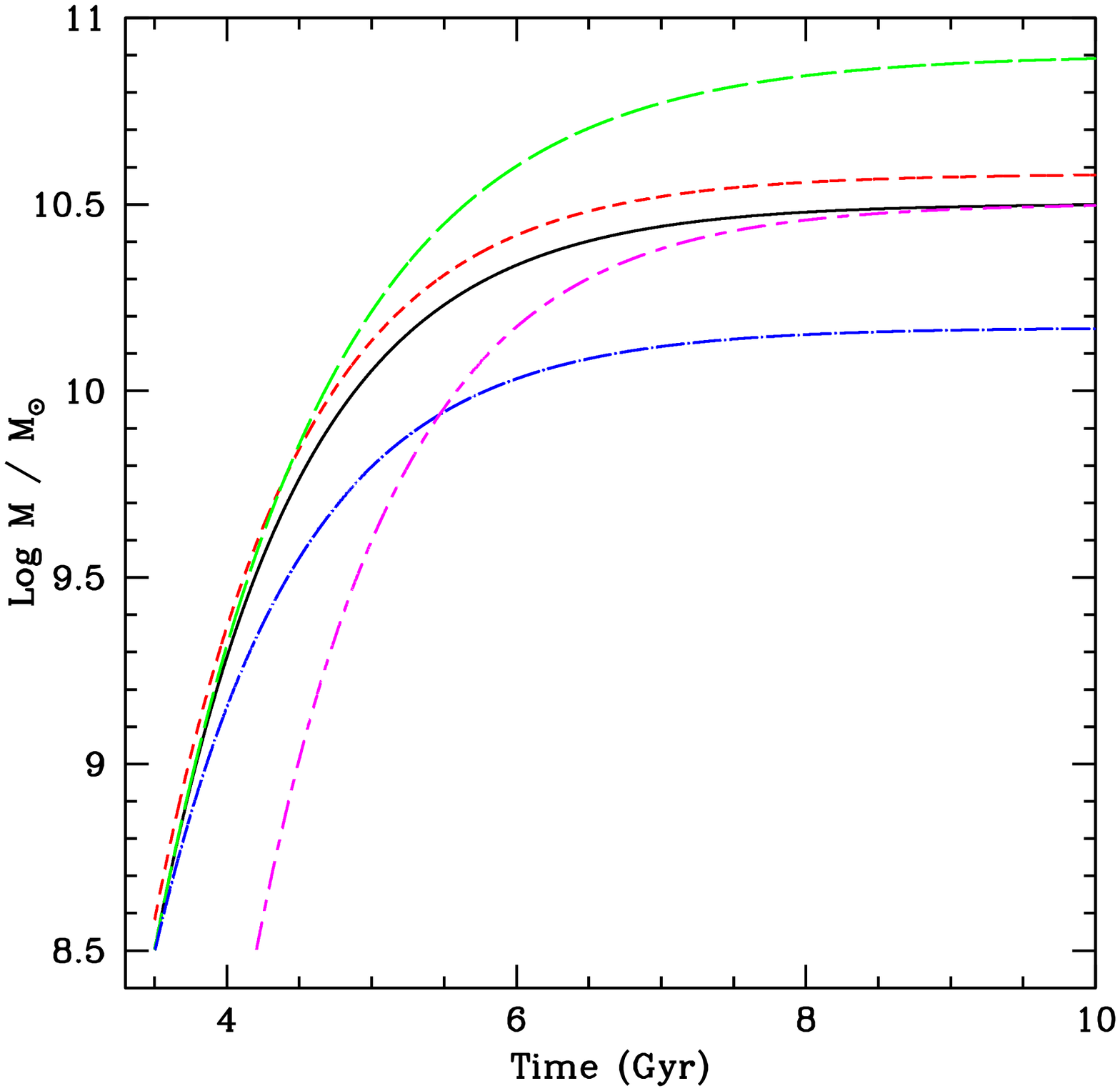}
\caption{Black: Evolution in the time-mass plane of an individual quasar from $10^{8.5} M_\odot$, starting at the Eddington luminosity and declining in Eddington ratio as $L/L_{Edd} = e^{-(t-t_0)/1.0}$, or $k_e = 1$ Gyr.  20\% increases in the initial mass (red), initial time (magenta), exponential decay time $k_e$ (green), and conversion factor $\kappa$ (blue) are also indicated.}
\label{fig:qsomodeltmexp}
\end{figure}

\begin{figure}
  \epsfxsize=3in\epsfbox{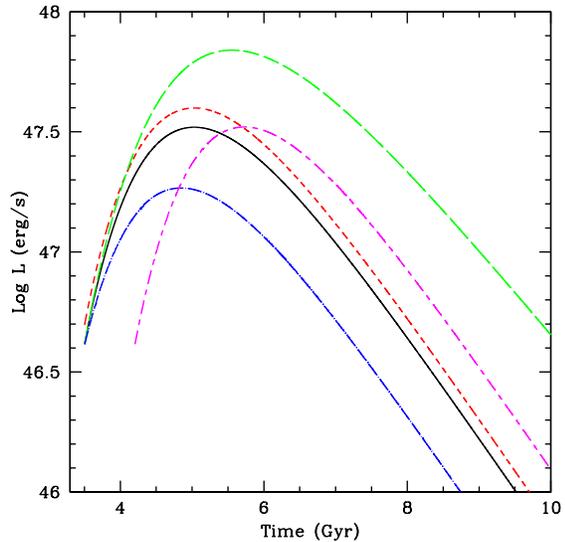}
\caption{Black: Evolution in the time-luminosity plane of an individual quasar from $10^{8.5} M_\odot$, starting at the Eddington luminosity and declining in Eddington ratio as $L/L_{Edd} = e^{-(t-t_0)/1.0}$, or $k_e = 1$ Gyr.  20\% increases in the initial mass (red), initial time (magenta), exponential decay time $k_e$ (green), and conversion factor $\kappa$ (blue) are also indicated.  Note that although the Eddington ratio is strongly decreasing, the increasing black hole mass results in a luminosity that first increases, then decreases.}
\label{fig:qsomodeltlexp}
\end{figure}

We can also consider the mass and luminosity evolution as a function of time.  In the resulting $t-M$ and $t-L$ planes (Figures \ref{fig:qsomodeltm} and \ref{fig:qsomodeltl}), changes in $k$ and $\kappa$ (again, green and blue, respectively) now produce slightly different shapes, and the final SMBH mass is sensitive to each of these parameters.  The exponential decay model exhibits sharper dependence upon track parameters in the $t-M$ plane but weaker dependent in the $M-L$ plane, as illustrated in Figures, \ref{fig:qsomodelmlexp} \ref{fig:qsomodeltmexp}, and \ref{fig:qsomodeltlexp}.

\section{Allowed Parameter Ranges}
\label{sec:parameters}

\begin{figure}
  \epsfxsize=3in\epsfbox{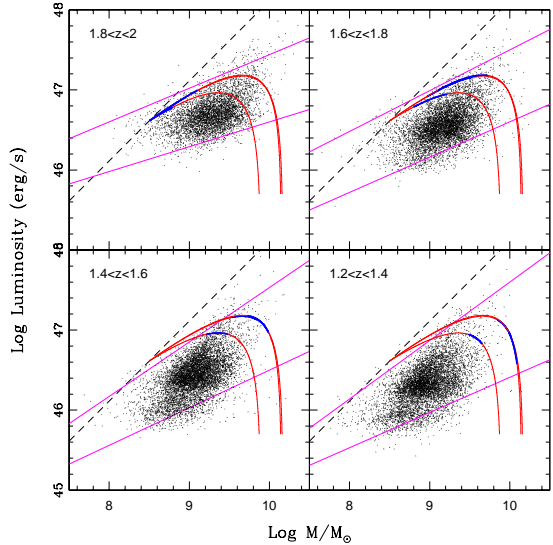}
\caption{The portions of two sample tracks (red) with $M_0 = 10^{8.5} M_\odot, t_0 = 3.5 \textrm{ Gyr}, k_e = 1.0 \textrm{ and } 0.5 \textrm{ Gyr}, \kappa = 0.65 \textrm{ and } 0.35$  lying in each redshift bin (blue) are constrained to lie within the observed quasar mass-luminosity locus (purple boundaries).}
\label{fig:trackexample}
\end{figure}

Using the two quasar track parametrizations from the previous section, we consider how individual quasars can evolve in mass and luminosity.  In this section, several methods are described for constraining the evolution of individual quasars by requiring that their track lie within the observed $M-L$ locus at every redshift, as in Figure \ref{fig:trackexample}.  The simplest method is to use the boundaries from Table \ref{table:boundaries} to restrict allowed mass and luminosity as a function of time.  We then proceed to consider further restrictions due to intrinsic quasar variability, methods for dealing with digitization problems when crossing from one redshift bin to the next in Table \ref{table:boundaries}, and methods for dealing with potential mismatches between the $M-L$ plane as calculated using H$\beta$-based and Mg{\small II}-based virial masses.

\begin{figure}
  \epsfxsize=3in\epsfbox{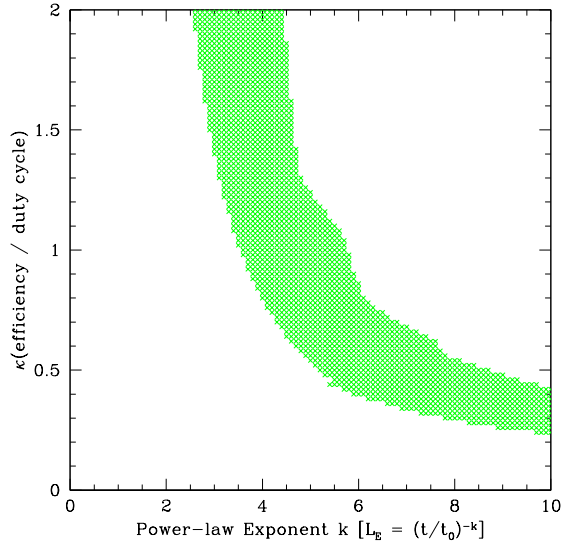}
\caption{Allowed (green) combinations of slope and conversion factor for quasars with a power-law-decaying Eddington ratio starting at $M_0 = 10^{8.5} M_\odot, t_0 = 3.5$ Gyr ($z=1.85$).}
\label{fig:8.5pl}
\end{figure}

\begin{figure}
  \epsfxsize=3in\epsfbox{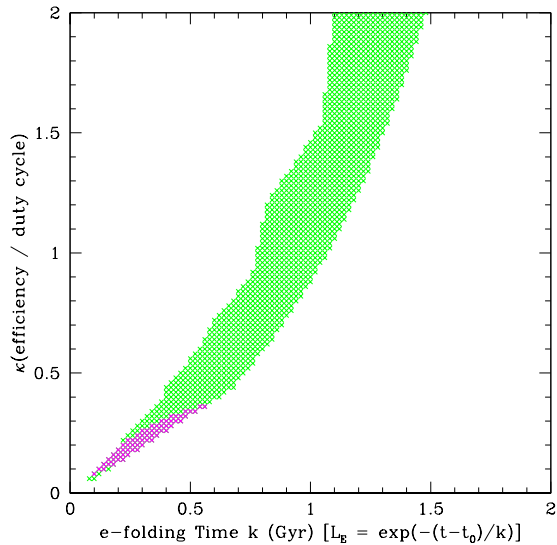}
\caption{Allowed (green, magenta) combinations of e-folding timescale and conversion factor for quasars with an exponentially-decaying Eddington ratio starting at $M_0 = 10^{8.5} M_\odot, t_0 = 3.5$ Gyr ($z=1.85$).  The magenta combinations correspond to quasars with a final mass of $9.5 < \log M/M_\odot < 9.6$.}
\label{fig:8.5exp}
\end{figure}

Using the most straightforward of these methods, directly applying the boundaries as in Table \ref{table:boundaries}, consider a quasar at $10^{8.5} M_\odot$ and $L_{Edd}$ at $1.8 < z < 2.0$, $t_0 = 3.5$ Gyr ($z = 1.85$ using cosmological parameters from \citet{WMAP3}).  Evolving this quasar forward, while constrained to lie within the Table \ref{table:boundaries} parameters, restricts the allowed slope $k$ and conversion factor $\kappa$ to narrow bands for both power-law (Figure \ref{fig:8.5pl}) and exponential (Figure \ref{fig:8.5exp}) decays.  The allowed parameters form roughly a one-dimensional locus.  
As indicated by the test parameters in \S~\ref{sec:tracks}, the turnoff time $t_f$ and final mass $M_f$ are sensitive to each of these parameters.  Requiring that the turnoff happen at a specific time or mass restricts the possibilities to a narrower set of possible parameters (e.g., $9.5 < \log M/M_\odot < 9.6$ in Figure \ref{fig:8.5exp}). 

Solutions with larger $\kappa$ correspond to SMBH that gain mass more slowly at a given luminosity and/or are more often quiescent, resulting in slower growth and a smaller mass at turnoff.  Solutions as $\kappa \rightarrow 0$ rapidly produce high-mass SMBH that then turns off.  These solutions therefore correspond to solutions in which growth is primarily non-luminous, and thus are disallowed by the requirement that most SMBH mass was acquired during a luminous quasar phase \citep{Soltan1982,Yu2002,Elvis2002}, and so are not plausible candidates for SMBH growth. 

\subsection{Effects of Variability}

Individual quasars in SDSS are observed to undergo $\sim 0.3$ dex variations in the optical on timescales of months to years \citep{structurefunction}.  The luminosities on the model tracks represent the characteristic luminosity of the quasar averaged over longer timescales.  However, the observed distribution of quasars includes a dispersion due to this intrinsic short-term variability.  the boundaries in Table \ref{table:boundaries} are similarly broadened, so the proper track must lie well within these boundaries.  Individual quasars might spend a small fraction of their time across a boundary (5\%, see \S~\ref{sec:tracks}).  

\begin{figure}
  \epsfxsize=3in\epsfbox{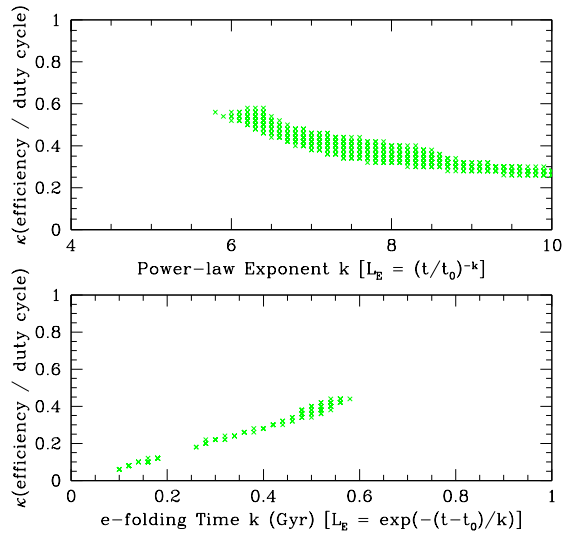}
\caption{Allowed combinations of e-folding timescale and conversion factor for quasars with Eddington ratio declining both as a power law and an exponential starting at $t_0 = 3.5$ Gyr and remaining at least $0.2$ dex from boundaries on the bright and faint ends of the quasar $M-L$ distribution until turnoff.  Initial masses are from $10^{8.13} - 10^{8.28} M_\odot$.}
\label{fig:8.2map}
\end{figure}

A restriction to lie $0.2$ dex away from each boundary produces no non-trivial solutions at $t_0 = 3.5, M_0 = 8.5$.  This might be because (1) the objects observed at $10^{8.5} M_\odot$ are, due to intrinsic variability, above their characteristic luminosity, (2) the redshift bins are too wide, or (3) quasar evolution does not lie on a simply-parametrized track of the sort we are fitting.  If $M_0$ is allowed to vary, keeping $t_0 = 3.5$ Gyr, solutions exist for the narrow range $8.13 < M_0 < 8.28$ (Figure \ref{fig:8.2map}).

\subsection{Common Tracks at all $M_0, t_0$}
\label{subsec:commonparameters}

\begin{figure}
  \epsfxsize=3in\epsfbox{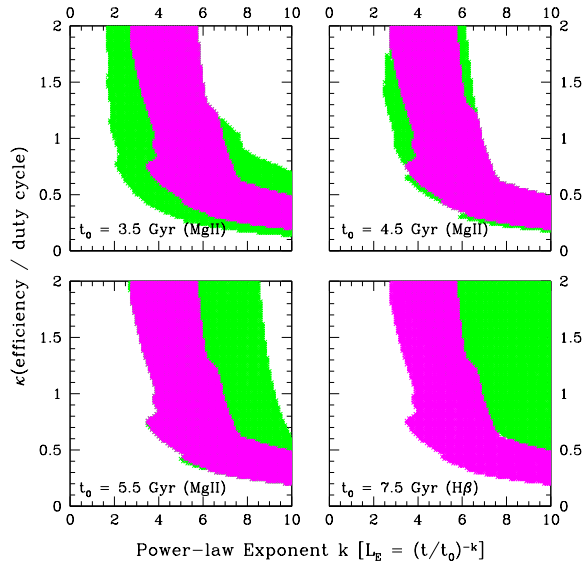}
\caption{Allowed combinations (green, magenta) of power-law slope $k$ and conversion factor $\kappa$ for quasars with Eddington ratio declining as a power law starting at four different choices of $t_0$.  The boundaries at each redshift are discontinuous, as given in Table \ref{table:boundaries}.  Discontinuities in the boundaries on allowed combination correspond to the quasar turnoff time moving from one redshift bin in Table \ref{table:boundaries} to another.  Magenta combinations of $k,\kappa$ are allowed in all four panels.}
\label{fig:allpl4panel}
\end{figure}

\begin{figure}
  \epsfxsize=3in\epsfbox{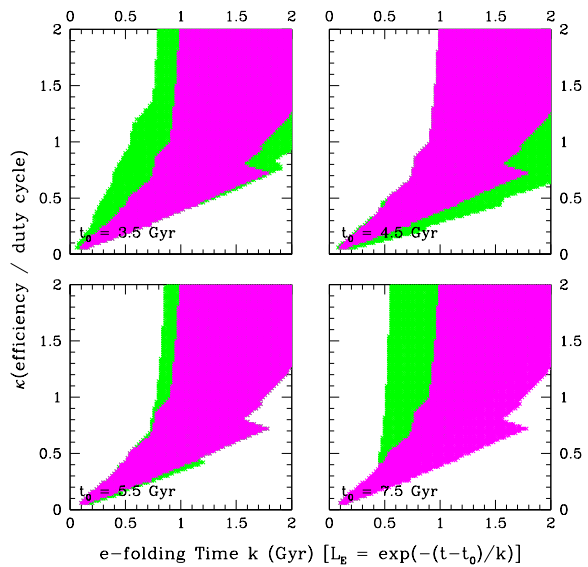}
\caption{Allowed combinations (green, magenta) of e-folding timescale $k$ and conversion factor $\kappa$ for quasars with an exponentially-decaying Eddington ratio starting at four different choices of $t_0$.  The boundaries at each redshift are given in Table \ref{table:boundaries}.  Discontinuities in the boundaries on allowed combination correspond to the quasar turnoff time moving from one redshift bin in Table \ref{table:boundaries} to another.  Magenta combinations of $k,\kappa$ are allowed in all four panels.}
\label{fig:allexp4panel}
\end{figure}

One natural possibility is that every SMBH grows with the same value $k$ and $\kappa$ but with different initial $t_0$ and $M_0$.  If so, this would be a strong indication that there is one, universal feedback process and that this simple model of quasar evolution has validity.  Figures \ref{fig:allpl4panel} and \ref{fig:allexp4panel} show the allowed values of $\kappa$ and $k$ for all power-law and exponential tracks, respectively, with any $M_0$ from $10^{5-10}$, departing from Eddington at $t_0 = 3.5, 4.5, 5.5,$ and $7.5$ Gyr.  Here, we restrict solutions to lie within each boundary at all times, and in \S~\ref{subsec:othermethods} will make these boundaries more restrictive due to quasar variability.

We can also restrict the above solutions to lie at least $0.2$ dex from each boundary to account for scatter due to intrinsic variability in the observed quasar population.

If all quasars follow a track with identical $k$ and $\kappa$, that track must be a valid solution in all panels of Figure \ref{fig:allpl4panel} or \ref{fig:allexp4panel}.  This would restrict possible combinations of $k$ and $\kappa$ to the ranges shown in magenta in Figures \ref{fig:allpl4panel} and \ref{fig:allexp4panel}.  There exist families of quasar tracks with the same $k$ and $\kappa$ valid at all times $t_0 > 3.5$ Gyr ($z < 1.91$).

In addition to those shown in Figures \ref{fig:allpl4panel} and \ref{fig:allexp4panel}, there are also solutions involving quasars with small ($M_0 \sim 10^6 M_\odot$) initial masses spending most of their lives at luminosities below the SDSS detection threshold and only becoming massive and luminous enough to lie within the SDSS catalog with $L \ll L_{Edd}$ and near $z=0$.  Although these solutions do lie within the boundaries in Table \ref{table:boundaries}, such solutions cannot describe the quasar distribution at $z \gg 0$, not do these solutions contribute to the Soltan argument \citep{Soltan1982}.  We have only included solutions in both this and the following sections (Figures \ref{fig:allpl4panel}--\ref{fig:allintbothMgoverlap}) that describe quasars which would lie in the SDSS catalog with $M_{BH} = M_0$ and redshift corresponding to $t_0$.  

\subsection{Other Interpolation Methods}
\label{subsec:othermethods}

\begin{figure}
  \epsfxsize=3in\epsfbox{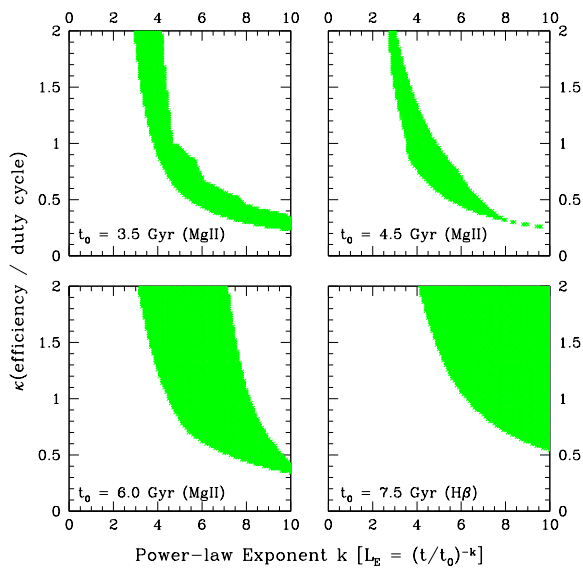}
\caption{Allowed combinations of e-folding timescale and conversion factor for quasars with Eddington ratio declining as a power law starting at four different choices of $t_0$.  The boundaries are interpolated from those in Table \ref{table:boundaries}.}
\label{fig:plint4panel}
\end{figure}

\begin{figure}
  \epsfxsize=3in\epsfbox{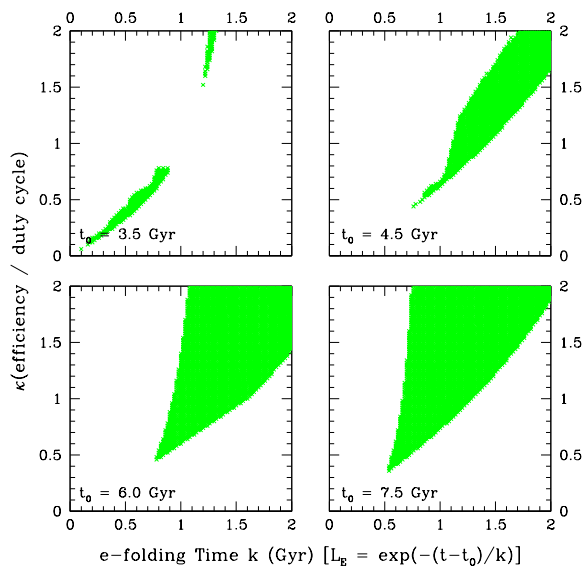}
\caption{Allowed combinations of e-folding timescale and conversion factor for quasars with an exponentially-decaying Eddington ratio starting at four different choices of $t_0$.  Boundaries on the quasar $M-L$ locus are interpolated from those in Table \ref{table:boundaries}.}
\label{fig:expint4panel}
\end{figure}

We can also restrict the above solutions to lie at least $0.2$ dex from each boundary to account for scatter due to intrinsic variability in the observed quasar population.  However, there are no non-trivial solutions lying at least $0.2$ dex away from each boundary with $t_0 = 4.5$ Gyr.  This may be because the boundaries change discontinuously when the quasar redshift moves across a bin boundary in Table \ref{table:boundaries}, and some boundaries in this table match poorly.  The greatest mismatches occur when virial masses transition from using Mg{\small II} to H$\beta$ broad emission lines at $z = 0.8$.   There are many ways one might correct for this binning problem.  As here we are simply developing models in an effort to establish rough guidelines for what properties are required of quasar evolution, we will just give the results of using a simple interpolation.  In Figures \ref{fig:plint4panel} and \ref{fig:expint4panel}, we take the boundaries in Table \ref{table:boundaries} to be correct at the midpoint (in time) of each bin, and interpolate in both slope and intercept to find the boundaries used at other redshifts.  Using these interpolated boundaries, Figures \ref{fig:plint4panel} and \ref{fig:expint4panel} include parameters for all tracks of the sort included in Figures \ref{fig:allpl4panel} and \ref{fig:allexp4panel}, with $10^5 < M_0 < 10^{10}$, departing from Eddington at the same four $t_0$. 

If quasar activity at all times is controlled by the identical physical constraints, all quasars might lie on a family of tracks with identical shape ($k$ and $\kappa$) but different $M_0$ and $t_0$.  There are no quasar tracks with either a power-law or exponential decline in Eddington ratio with $k$ and $\kappa$ lying in the allowed range in all panels of either Figure.  Thus, families of tracks with the same $k,\kappa$ at all times are disallowed with this interpolation.

\subsection{Restriction to MgII masses}

\begin{figure}
  \epsfxsize=3in\epsfbox{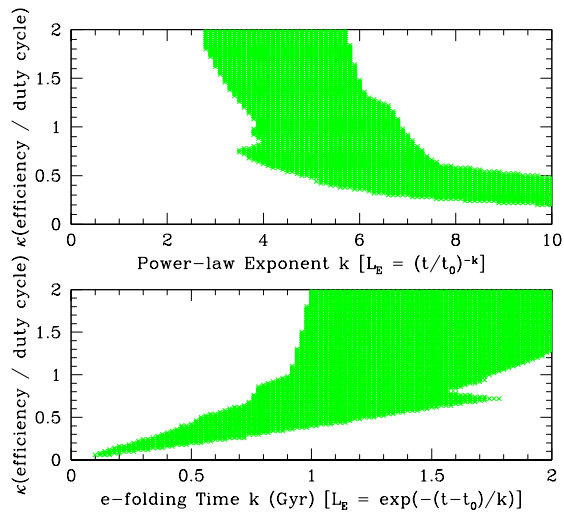}
\caption{Quasar track parameters allowed for the three choices of $t_0 < 6.0$ Gyr (with Mg{\small II}-derived masses) in Figures \ref{fig:allpl4panel} and \ref{fig:allexp4panel}.  The boundaries at each redshift are given in Table \ref{table:boundaries}.}
\label{fig:allbothMgoverlap}
\end{figure}

\begin{figure}
  \epsfxsize=3in\epsfbox{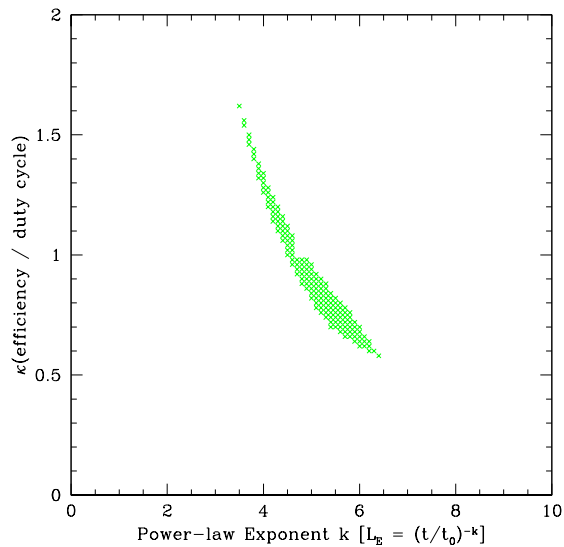}
\caption{Quasar track parameters allowed for the three choices of $t_0 < 6.0$ Gyr (with Mg{\small II}-derived masses) in Figures \ref{fig:plint4panel} and \ref{fig:expint4panel}.  There are no tracks allowed at all three choices of $t_0$ for an exponential decline in quasar Eddington ratio.  The boundaries at each redshift are interpolated from those in Table \ref{table:boundaries}.}
\label{fig:allintbothMgoverlap}
\end{figure}

Finally, recent studies comparing H$\beta$- and Mg{\small II}-based virial masses have proposed that a substantial correction might be required to bring the two into agreement \citep{Onken2008,Risaliti2009}, although the correction might not be needed for the brightest quasars \citep{Steinhardt2010c}.  The allowed parameters in the three panels with $t_0 < 6.0$ Gyr of Figures \ref{fig:allpl4panel} and \ref{fig:allexp4panel} are primarily determined using Mg{\small II} masses, while the $t_0 = 7.5$ Gyr ($z = 0.68$) panel is determined using primarily H$\beta$ masses.  Requiring that parameters appear only in all three Mg{\small II} panels produces the allowed parameters in Figures \ref{fig:allbothMgoverlap} and, for interpolated boundary conditions, the power-law parameters indicated in Fig. \ref{fig:allintbothMgoverlap}.  There are still no parameters for an exponential decay that appear in all three Mg{\small II} panels of Figure \ref{fig:allexp4panel} (for the interpolated boundary conditions).
 
\section{Discussion}
\label{sec:trdiscussion}

We have shown that the quasar mass-luminosity plane can be used to restrict possible tracks for the evolution of individual quasars.  The allowed parameters presented in this paper correspond to simple models and analysis methods.  In \S~\ref{sec:trdiscussion}, we discuss some of the properties that appear to be common to all allowed tracks.  Additional tuning of the quasar track shape, e.g., requiring a different minimum distance from boundaries on the quasar distribution, or the use of interpolation methods produces altered sets of allowed parameters.  These figures are intended to serve only as a rough guide to the types of changes that result from using different assumptions.

The production of these tracks was motivated by a consideration of the surprising synchronization of quasar luminosity evolution at fixed black hole mass.  It was hoped that the seeming independence of quasar accretion rates on host galaxy parameters might allow quasar observations at different redshifts to be matched as lying along a common evolutionary track.  While it may have seemed highly unlikely that the many complications provided by galactic dynamics, merger history, and asynchronous quasar behavior were all truly negligible, surprisingly we find that simple, parametrized models indeed may provide a good description of the evolution of individual quasars.  Further, it may even be possible that one family of similar tracks can describe the evolution of all Type 1 quasars.

\subsection{Tracks}

We have tested models with both an exponential and a power-law decline in quasar luminosity when constrained to pass through the observed quasar locus at all lower $z$.  The power-law decline model has been proposed by \citet{Hopkins2009}.  The quasar tracks are sensitive to $\sim 20\%$ parameter changes (Figures \ref{fig:qsomodelml} and \ref{fig:qsomodeltm}), showing that these tracks provide a sensitive probe of the details of quasar fueling and feedback.  

While the allowed ranges are sensitive to the details of how the boundaries of the quasar distribution are defined at each redshift, the allowed tracks have some characteristic general properties:
\begin{enumerate}
\item{Typically, tracks are allowed over a restrictive range of rates of Eddington ratio decline and ``accretion conversion factors'' $\kappa$.  Different combinations of decline rate $k$ and $\kappa$ result in different amounts of luminous mass growth from the start of the quasar phase until turnoff, as shown in Figure \ref{fig:qsomodeltm}.}
\item{All allowed tracks are characterized by a steep decline in Eddington ratio.  For a power-law decline $L/L_{Edd} \propto t^{-k_e}$, even allowing $\kappa$ to be as large as 2 results in a track with $k_e > 4$, while placing $\kappa \leq 0.5$ as expected would result in declines with $k_e > 6$ or steeper.}
\end{enumerate}

\subsection{Rapid Decline}

We might have expected that only densities of gas and dust within the host galaxy density should be important, but the synchronization of the quasar accretion rates \citep{Steinhardt2010b} implies that this might not be the case.  In a matter-dominated universe (as existed at $z = 2$), the matter density falls as $(1+z)^{-3} \propto t^{-2}$.  If quasar fuelling were linked to densities in the intergalactic medium, then, the falloff could not be $t^{-6}$ or steeper.  The rate of random two body collisions, e.g., galaxy collisions is proportional to the square of their number density in the universe, giving $t^{-4}$.  A dependence of $t^{-6}$ or steeper could arise from collisions involving at least three bodies.  Alternatively, other astrophysics might dictate a steeper decline, as proposed by \citet{Peng2010} for ``quenching'' processes dominating galactic evolution.  In the exponential case $L/L_{Edd} \propto e^{-k_e}$, the decline must also be quite sharp, typically with an e-folding time of 1 Gyr or shorter.  

Both sets of allowed tracks present quasars as relatively short-lived objects, living only $\sim 1-2$ Gyr from their arrival above SDSS detection near the Eddington ratio until turnoff.  The SDSS includes a large quasar population on $0.2 < z < 4.1$, a range spanning 10 Gyr.  If the typical quasar can indeed only exist in the SDSS catalog for $\sim 1-2$ Gyr, then a typical galaxy can only contain a quasar a maximum of $\sim 10-20\%$ of that time, even if the quasar duty cycle includes no periods of quiescence.  A mostly-quiescent duty cycle would further reduce the observed fraction of galaxies containing quasars.  Since this observed fraction is $\sim 10-20\%$ near $z=2$ and only slightly lower towards the edges of the SDSS quasar redshift range (cf. \citet{Martini2009}), we conclude that the typical quasar indeed has a duty cycle such that it is luminous closer to 100\% than to 10-20\% of the time.  For the duty cycle to fully reach 100\%, a track with either a large ($\sim 25\%$ or more) radiative efficiency or a steep decline (towards $t^{-10}$) appears to be required.  The implication is that SMBH evolution might not involve ``flickering'', i.e., interspersed periods of luminous accretion and quiescence on a $\sim 10^7$ year timescale \citep{Hatziminaoglou2001}.  

Instead, a simpler picture suggests itself: (1) the SMBH is seeded; (2) the SMBH grows until it enters the SDSS catalog at the low-mass end and at the Eddington luminosity; (3) the SMBH accretes as a Type 1 quasar for $1-2$ Gyr while the Eddington ratio declines sharply; and (4) the SMBH permanently ceases its rapid, unobscured, luminous accretion, with various possible post-turnoff states including quiescence, Seyferts, and Type 2 quasars.  This picture also requires a low value of $\kappa$, corresponding to decline proportional to $t^{-6}$ or steeper.

The possibility of a short, $1-2$ Gyr quasar lifetime is particularly intriguing in light of two additional results from the $M-L$ plane.  The characteristic luminosity for quasar accretion at fixed mass and redshift requires that accretion rates be synchronized to within $1-2$ Gyr \citep{Steinhardt2010b}.  Also, quasar turnoff is synchronized, depending upon the mass, to within $0.75-3$ Gyr for $M_BH > 10^9 M_\odot$ \citep{Steinhardt2010b}.  Perhaps the similarity of these three synchronization timescales could be explained by a synchronization in the times with which quasars turn and follow a common track in the $M-L$ plane combined with short lifetimes.

A short-lived Type 1 quasar phase might seem to violate the Soltan argument because the SMBH spends most of its time in another state.  The Soltan argument shows that most of the total quasar mass in the universe was accreted luminously in Type 1 quasar states.  However, the Soltan argument only places a limit on the last 2-3 e-foldings of mass growth, during which most of the mass is added.  Prior to these last e-foldings, we do not know how much of the growth takes place through luminous accretion.  Even these short tracks with decline between $t^{-6}$ and $t^{-10}$ grow the SMBH by 1-1.4 dex, i.e., 2-3 e-foldings, so such solutions are allowed. 

\subsection{A Common Track}

As shown in Figures Figures \ref{fig:allpl4panel},\ref{fig:allexp4panel}, \ref{fig:allbothMgoverlap}, and \ref{fig:allintbothMgoverlap}, it is possible that one scaling law for feedback with universal parameters might be able to describe the evolution of all quasars at all initial masses and times.  The existence of a characteristic luminosity at each combination of mass and redshift is insufficient to require such a uniformity among the evolution of individual quasars.  However, (1) the synchronization between quasars at fixed mass, (2) the narrow luminosity range at fixed mass and redshift, and (3) the sharp peak in number density at a single, characteristic luminosity at fixed mass and redshift (cf. \citet{Steinhardt2010b}) make such a model intriguing.  

\section{Conclusions}

In this paper, we have investigated tracks for SMBH accretion histories and have shown that the quasar mass-luminosity plane constrains these models remarkably precisely.  The most intriguing result is that we can rule out models in which the SMBH accretion rate is proportional to the matter number density $n$ in the universe.  Even models in which the accretion rate is proportional to $n^2$ are not allowed without a combination of higher-than-expected radiative efficiency and a SMBH that spends most of its time quiescent, rather than in a quasar state.

This paper is an intermediate phenomenological step that has produced constraints that seem to be required of theoretical models for SMBH growth.  The next step is to produce a physical model of fuelling and feedback leading to quasar tracks satisfying these constraints.

The authors would like to thank Lars Hernquist and Norm Murray for valuable comments.  This work was supported in part by Chandra grant number G07-8136A.  This work was supported by World Premier International Research Center Initiative (WPI Initiative), MEXT, Japan.

\bibliographystyle{mn2e}
\bibliography{ms}

\end{document}